\documentstyle{article}

\advance \textheight by \topskip
\def\yes{y } 
\ifx\ifig\yes
  \input epsf
  
  \def\INSERTFIG#1#2#3{\epsfxsize=#1in \nopagebreak \begin{center} \hbox
    to\hsize{\hfil\boxit{\epsffile{#2}}\hfil} {\sl #3} \end{center}}
\else
  \def\INSERTFIG#1#2#3{\par\noindent #3\par}
\fi

\begin{document}

\begin{titlepage}
\title{The Causal Interpretation of Dust and Radiation Fluids Non-Singular  Quantum Cosmologies}
\author{{\bf J. Acacio de Barros}\thanks{e-mail address: 
acacio@fisica.ufjf.br} ${}^{a}$\\
{\bf N. Pinto-Neto}\thanks{e-mail address:
nen@lca1.drp.cbpf.br}${}^{\;\;b}$\\ and \\
{\bf M. A. Sagioro-Leal} ${}^{a}$\\
${}^{a}${\it Departamento de F\'{\i}sica - ICE}\\
{\it Universidade Federal de Juiz de Fora}\\
{\it 36036-330, Juiz de Fora, MG, Brazil}\\
${}^{b}${\it Centro Brasileiro de Pesquisas F\'{\i}sicas/Lafex}\\
{\it Rua Xavier Sigaud, 150 - Urca}\\
{\it 22290-180, Rio de Janeiro, RJ,Brazil}}
\maketitle
\begin{abstract}
We apply the causal interpretation of quantum mechanics to homogeneous
and isotropic quantum cosmology where the sources of the gravitational
field are either dust or radiation perfect fluids. We find non-singular
quantum trajectories which tends to the classical one when the scale
factor becomes much larger then the Planck length. In this situation,
the quantum potential becomes negligible. There are no particle
horizons.  As radiation is a good approximation for the matter content
of the early universe, this result suggests that the universe can be
eternal due to quantum effects.
\vspace{0.7cm}

PACS number(s): 98.80.H, 03.65.Bz
\end{abstract}
\end{titlepage}

\section{Introduction}

The appearance of initial singularities in the classical cosmological models
which better describe the universe we live in constitutes a big puzzle to
all cosmologists. Until now, singularities are out of the scope of any
physical theory. If we assume that a physical theory can describe the whole
Universe at every instant, even at its possible moment of creation (which is
the best attitude because it is the only way to seek the limits of physical
science), then these classical singular points must be avoided. Indeed, no
one expects that classical general relativity continues to be valid under
extreme situations of very high energy density and curvature. In particular,
it is very plausible that quantum gravitational effects become important
under these conditions, eliminating the singularities that appear
classically. To see if this is indeed the case, we should construct a theory
of quantum cosmology. However, any quantum theory when applied to cosmology
presents new profound conceptual problems. How can we apply the standard
probabilistic Copenhaguen interpretation to a single system as the
Universe? Where in a quantum Universe can we find a classical domain where
we could construct our classical measuring apparatus to test and give sense
to the quantum theory? Who are the observers of the whole Universe? This is
not a problem of quantum gravity alone, because there is no problem with the
concept of an ensemble of black holes and a classical domain outside it, but
it is specific of quantum cosmology. As we cannot apply the Copenhaguen
interpretation to quantum cosmology, we will adopt an alternative
non-probabilistic interpretation, which circumvents the measurement problem
because it is an ontological interpretation of quantum mechanics: it is not
necessary to have a measuring apparatus or a classical domain in order to
recover physical reality. It is the causal
interpretation of quantum mechanics \cite{boh,hol}.

In this letter, we apply the causal interpretation of quantum mechanics to
some specific quantum states already found in the literature \cite
{gotay,lemos}, and show that the Bohmian (quantum) trajectories are
non-singular, tending to the classical trajectories when the scale factor is
large. The quantum potential is the responsible for this behavior. When the
scale factor is small, the quantum potential becomes large creating an
effective repulsive force avoiding the singularity. When the scale factor is
large, the quantum potential becomes negligible and the classical potential
dominates. The minisuperspace models we study are constituted of a
Friedman-Robertson-Walker (FRW) metric with either dust or radiation perfect
fluids \cite{gotay,lemos}. The ADM quantization procedure is performed
because in these cases there is a preferable time variable which rends the
quantum equations into a very simple Shroedinger form. The quantum solutions
obtained are gaussians \cite{gotay,lemos}. The dust case is rather academic
but the radiation case is important because it can mimic quite well the
matter content at the very early universe. In all cases we obtain
non-singular eternal models without any particle horizon.

This letter is organized as follows. In the next section we make a summary
of the causal interpretation and its application to quantum cosmology. In
section 3 we abridge the results of Refs. \cite{gotay,lemos} which concern
this letter. In section 4 we apply the causal interpretation to the
solutions presented in section 3, obtaining the Bohmian trajectories. We
compare our results with those of Refs. \cite{gotay,lemos}. We end with
comments and conclusions.

\section{The causal interpretation of quantum mechanics}

In this section, we will review the causal interpretation of quantum
mechanics. Let us begin with the Schr\"{o}dinger equation, in the coordinate
representation, for a non-relativistic particle with the hamiltonian $H =
p^2 / 2m + V(x)$:

\begin{equation}  \label{bsc}
i \hbar \frac{d \Psi (x,t)}{d t} = [-\frac{\hbar ^2}{2m} \nabla ^2 + V(x)]
\Psi (x,t) .
\end{equation}
Writing $\Psi = R \exp (iS/\hbar)$, and substituting it into (\ref{bsc}), we
obtain the following equations: 
\begin{equation}  \label{bqp}
\frac{\partial S}{\partial t}+\frac{(\nabla S)^2}{2m} + V -\frac{\hbar ^2}{2m%
}\frac{\nabla ^2 R}{R} = 0 ,
\end{equation}
\begin{equation}  \label{bpr}
\frac{\partial R^2}{\partial t}+\nabla .(R^2 \frac{\nabla S}{m}) = 0 .
\end{equation}

The usual probabilistic interpretation takes equation (\ref{bpr}) and
understands it as a continuity equation for the probability density $R^2$
for finding the particle at position $x$ and time $t$ if a measurement is performed. All physical
information about the system is contained in $R^2$, and the total phase $S$
of the wave function is completely irrelevant. In this interpretation,
nothing is said about $S$ and its evolution equation (\ref{bqp}). However,
examining equation (\ref{bpr}), we can see that $\nabla S /m$ may be
interpreted as a velocity field, suggesting the identification $p=\nabla S$.
Hence, we can look to equation (\ref{bqp}) as a Hamilton-Jacobi equation for
the particle with the extra potential term $-\hbar ^2 \nabla ^2 R /2m R$.

After this preliminary, let us introduce the causal interpretation of
quantum mechanics, which is based on the {\it two} equations (\ref{bqp}) and
(\ref{bpr}), and not only in the last one as is the Copenhaguen
interpretation:

i) A quantum system is composed of a particle {\it and} a field $\Psi$
(obeying the Schr\"{o}dinger equation (\ref{bsc})), each one having its own
physical reality.

ii) The quantum particles follow trajectories $x(t)$, {\it independent on
observations}. Hence, in this interpretation, we can talk about trajectories
of quantum particles, contrary to the Copenhaguen interpretation, where only
positions at one instant of time have a physical meaning.

iii) The momentum of the particle is $p=\nabla S$.

iv) For a statistical ensemble of particles in the same quantum field
$\Psi$ the probability density is $P=R^2$. Equation (\ref{bpr})
guarantees the conservation of $P$.

Let us make some comments:

a) Equation (\ref{bqp}) can now be interpreted as a Hamilton-Jacobi type
equation for a particle submited to an external potential which is the
classical potential plus a new quantum potential 
\begin{equation}  \label{qp}
Q \equiv -\frac{\hbar ^2}{2m}\frac{\nabla ^2 R}{R} .
\end{equation}
Hence, the particle trajectory $x(t)$ satisfies the equation of motion 
\begin{equation}  \label{beqm}
m \frac{d^2 x}{d t^2} = -\nabla V - \nabla Q .
\end{equation}

b) Even in the regions where $\Psi$ is very small, the quantum potential can
be very high, as we can see from equation (\ref{qp}). It depends only on the
form of $\Psi$, not on its absolute value. This fact brings home the
non-local and contextual character of the quantum potential\footnote{%
This fact becomes evident when we generalize the causal interpretation to a
many particles system.}. This is very important because Bell's inequalities
together with Aspect's experiments show that, in general, a quantum theory
must be either non-local or non-ontological. As Bohm's interpretation is
ontological, it must be non-local, as it is. The quantum potential is
responsible for the quantum effects.

c) This interpretation can be applied to a single particle. In this case,
equation (\ref{bpr}) is just an equation to determine the function $R$,
which forms the quantum potential acting on the particle via equation (\ref
{beqm}). It is not necessary to interpret $R^2$ as a probability density and
hence it may not be normalizable. The interpretation of $R^2$ as a probability
density is appropriate only in the case mentioned in item (iv) above. The
causal interpretation is not, in essence, a probabilistic interpretation.

d) The classical limit is very simple: we only have to find the conditions
for having $Q=0$.

e) There is no need to have a classical domain because this interpretation
is ontological. The question on why in a real measurement we do not see
superpositions of the pointer apparatus is answered by noting that, in a
measurement, the wave function is a superposition of non-overlaping wave
functions. The particle will enter in one region, and it will be influenced
by the unique quantum potential obtained from the sole non-zero wave
function defined on this region. The particle cannot jump to other branchs
because it cannot pass through nodal points of the wave function.

In the next section we will perform an ADM quantization of
Friedman-Robertson-Walker minisuperspace models containing either dust or
radiation perfect fluids. In these cases, a prefered time variable can be
selected in terms of potentials of the velocity field of the fluids. As a
time variable will be chosen before quantization, the quantum equations will
be exactly like the Schr\"{o}dinger equation. Instead of the particle
position, as in the above example, the single degree of freedom of these
quantum models will be the scale factor of the universe. The interpretation
of the quantum solutions will run analogously to what was described above.
The time evolution of the scale factor will be different from the classical
one due to the presence of an extra quantum potential term in the modified
Hamilton-Jacobi equation it satisfies.

\section{The ADM quantization of dust and radiation minisuperspace models}

In this section we present the minisuperspace models which we will analyze
using the causal interpretation. These models are obtained from the
quantization of the dust and radiation fluids through the ADM prescription.

For the quantization of the radiation filled FRW model we follow Ref. \cite
{lemos}. We start with the line element 
\[
ds^{2}=-N(t)^{2}dt^{2}+a^{2}(t)\sigma _{ij}dx^{i}dx^{j}. 
\]
where $\sigma _{ij}$ is the metric of constant curvature three-surfaces. The
full action for a perfect fluid is given by 
\begin{equation}
S=\int_{M}d^{4}x\sqrt{-g}(R^{(4)}+p)+2\int_{\partial M}d^{3}x\sqrt{h}%
h_{ij}K^{ij},  \label{actionnivaldo}
\end{equation}
where $p$ is the pressure, $R^{(4)}$ is the scalar curvature, $h_{ij}$ is
the 3-metric on $\partial M,$ $K^{ij}$ is the second fundamental form of the
boundary, and we choose $c=16\pi G=1$.

The action (\ref{actionnivaldo}) can be reduced to 
\[
S_{r}=\int dt\left[ \dot{a}\,p_{a}-\dot{\varphi}p_{\varphi }+\dot{S}p_{S}-N%
{\cal H}\right] , 
\]
where we are using Schutz's fluid variables $\varphi ,$ $\lambda ,$ $\gamma ,
$ $\theta ,$ $S$ \cite{Schutz}, in terms of which the four-velocity of the
fluid is written as 
\begin{equation}  \label{veloc}
U_{\nu}=\frac{1}{\mu}(\partial _{\nu} \varphi + \lambda \partial _{\nu}
\gamma + \theta \partial _{\nu} S).
\end{equation}
The quantity $\mu$ is the specific enthalpy. The super-Hamiltonian ${\cal H}$
takes the form 
\[
{\cal H}=-\left( \frac{p_{a}^{2}}{24a}+6ka\right) +p_{\varphi
}^{4/3}a^{-3}e^{S}. 
\]
For details, see Refs. \cite{russos,tipler}.

Performing an ADM reduction using the conformal-time gauge $N=R$ in the
radiation case, the reduced action becomes \cite{tipler} 
\[
S_{r}=\int dt\left[ \dot{a}\,p_{a}-\left( \frac{p_{a}^{2}}{24}%
+6ka^{2}\right) \right] , 
\]
where $k=+1,$ $0$ or $-1$, for spherical, flat, or hyperbolic spacelike
sections of the three space with metric $\sigma _{ij}$, respectively. The
Hamiltonian in the reduced phase space, which has only $a$ as a degree of
freedom, takes the very simple form 
\[
H=\frac{p_{a}^{2}}{24}+6ka^{2}. 
\]
The classical solutions are: 
\begin{equation}
a = a_0 \left\{ 
\begin{array}{ll}
\sin t, & \quad \mbox{for } k=1 \\ 
t, & \quad \mbox{for } k=0 \\ 
\sinh t, & \quad \mbox{for } k=-1
\end{array}
.\right.  \label{classical}
\end{equation}
which are the well known solutions for radiation in conformal time.

The quantized Hamiltonian, in units where $\hbar =1,$ is 
\begin{equation}
\hat{H}=-\frac{1}{24}\frac{d^{2}}{da^{2}}+6ka^{2}.  \label{ho}
\end{equation}
As $a\geq 0$, the requirement that $\hat{H}$ must be self-adjoint leads to
the restriction 
\begin{equation}
\psi ^{\prime }(0)=\alpha \psi (0)  \label{boundary}
\end{equation}
on the wave function $\psi ,$ where $\alpha $ is a parameter in the interval 
$(-\infty ,\infty ].$ Two solutions will be obtained, one for $\alpha =0$
and the other for $\alpha =\infty .$

The propagators for the Hamiltonian (\ref{ho}) are well known. However, we
have the extra constraint that $a\geq 0.$ For this reason, the Hilbert space
is restricted to functions in $L^{2}(0,\infty ).$ For the case where $\alpha
=0$ the propagator is 
\begin{equation}
G^{(I)}(a,a^{\prime },t)=G(a,a^{\prime },t)+G(a,-a^{\prime },t),
\label{propagatora}
\end{equation}
where $G(a,a^{\prime },t)$ is the usual harmonic oscillator propagator for a
system with mass $m=12$ and angular frequency $\omega =\sqrt{k}.$ Seting $%
\alpha = 0$ in Eq. (\ref{boundary}), we take a gaussian wavepacket as the
initial normalized state, 
\begin{equation}
\psi _{0}^{(I)}(a)=\left( \frac{8b}{\pi }\right) ^{1/4}\exp (-\beta a^{2}), 
\label{gaussian0}
\end{equation}
where $\beta =b + iB,$ $B$ and $b$ are real, and $b>0.$ The wave function at
time $t$ given by Eqs. (\ref{propagatora}) and (\ref{gaussian0}) is 
\begin{eqnarray}
\psi ^{(I)}(a,t) &=&\left( \frac{8b}{\pi }\right) ^{1/4}\left[ \frac{6\sqrt{k%
}}{\cos (\sqrt{k}t)[\beta \tan (\sqrt{k}t)-6i\sqrt{k}]}\right] ^{1/2}
\label{psia} \\
&&\times \exp \left\{ \frac{6i\sqrt{k}}{\tan (\sqrt{k}t)}\left( 1+\frac{6i%
\sqrt{k}}{\cos ^{2}(\sqrt{k}t)[\beta \tan (\sqrt{k}t)-6i\sqrt{k}]}\right)
a^{2}\right\} .  \nonumber
\end{eqnarray}
The expectation value for the scale factor $a$ can be computed from (\ref
{psia}), and is 
\begin{equation}
\langle \hat{a}\rangle _{t}^{(I)}=\frac{1}{12}\sqrt{\frac{2}{\pi b}}\left\{ 
\begin{array}{ll}
\sqrt{b^{2}\sin ^{2}t+(6-B\tan t)^{2}\cos ^{2}t} & \quad \mbox{for }k=1 \\ 
\sqrt{b^{2}t^{2}+(6-Bt)^{2}} & \quad \mbox{for }k=0 \\ 
\sqrt{b^{2}\sinh ^{2}t+(6-B\tanh t)^{2}\cosh ^{2}t} & \quad \mbox{for }k=-1
\end{array}
.\right.  \label{expectationa}
\end{equation}

For $\alpha =\infty $ the propagator is 
\begin{equation}
G^{(II)}(a,a^{\prime },t)=G(a,a^{\prime },t)-G(a,-a^{\prime },t).
\label{propagatorb}
\end{equation}
We take as the initial state the wave packet 
\begin{equation}
\psi _{0}^{(II)}(R)=\left( \frac{8b}{\pi }\right) ^{1/4}R\exp (-\beta R^{2}).
\label{gaussiannot}
\end{equation}
The evolution of (\ref{gaussiannot}) governed by the propagator (\ref
{propagatorb}) is 
\begin{eqnarray}
\psi ^{(II)}(a,t) &=&\left( \frac{128 b^{3}}{\pi }\right) ^{1/4}\left[ \frac{%
216ik^{3/2}}{\sin ^{3}(\sqrt{k}t)}\right] \left[ \beta -\frac{6i\sqrt{k}}{%
\tan (\sqrt{k}t)}\right] ^{-3/2}  \label{psib} \\
&&\times R\exp \left\{ \frac{6i\sqrt{k}}{\tan (\sqrt{k}t)}\left( 1+\frac{6i%
\sqrt{k}}{\cos ^{2}(\sqrt{k}t)[\beta \tan (\sqrt{k}t)-6i\sqrt{k}]}\right)
R^{2}\right\} .  \nonumber
\end{eqnarray}
The expectation value for the scale factor $a$ with the wavefunction (\ref
{psib}) is 
\begin{equation}
\langle \hat{a}\rangle _{t}^{(II)}=2\langle \hat{a}\rangle _{t}^{(I)}.
\label{expectationb}
\end{equation}

We now turn our attention to the dust filled minisuperspace model presented
by Gotay and Demaret \cite{gotay}. Once again Schutz's variables and a FRW
metric are used. In this case the field $\varphi$ in Eq. (\ref{veloc}) is
the only velocity potential which is non null. The time to be chosen will be
the proper time of the dust particles defined by $\varphi=-t$, which is
equivalent to choose $N=1$. Seting $p=0$ and $\mu=1$, the ADM reduced
super-Hamiltonian becomes (see Ref. \cite{gotay}) 
\[
{\cal H}=-\left( \frac{p_{a}^{2}}{24a}+6ka\right) - p_{\varphi}. 
\]
The reduced Hamiltonian, with the above choice of time, is 
\[
H(a,p_a) = \frac{p_{a}^{2}}{24a}+6ka. 
\]
If we perform the canonical transformation 
\[
x=\frac{4}{3}\sqrt{6} a^{3/2},\qquad p_{x}=\frac{\sqrt{6}}{12} a^{-1/2}
p_{a} 
\]
the Hamiltonian takes the simple form 
\[
H(x,p_{x})=p_{x}^{2}+Kx^{2/3}, 
\]
where $K=\frac{3}{2}6^{3/2}k.$ Taking $k=0$, the classical solution is
simple to obtain. It is $x(t)=t$ or equivalently $a(t)\propto t^{2/3}$. The
quantized Hamiltonian for $k=0$ is, in units where $\hbar =1,$ 
\[
\hat{H}=-\frac{d^{2}}{dx^{2}}. 
\]
Once again, the requirement of self-adjointness of $\hat{H}$ yields the
boundary conditions $\psi ^{\prime }(0)=\alpha \psi (0)$. Choosing $\alpha =0
$, we can evolve the initial gaussian wave function 
\begin{equation}
\psi _{0}(x)=\left[ \frac{8b}{\pi }\right] ^{1/4}\exp (-\beta x^{2}),
\label{gauss2}
\end{equation}
by applying the corresponding propagator, analogously to what was done in
the radiation case. As before, $\beta=b+iB$ with $b\geq 0$. The wave
function at time $t$ is given by: 
\begin{equation}
\psi (x,t)=\left[ \frac{8b}{\pi }\right] ^{1/4}(1+4i\beta t)^{-1/2}\exp
\left[ \frac{-\beta x^{2}}{1+4i\beta t}\right] .  \label{psic}
\end{equation}
The expectation value for $x$ can easily be computed, and is 
\begin{equation}
\langle \hat{x}\rangle _{t}=\left( \frac{1}{2\pi b}\right) ^{1/2}\left[
16b^{2}t^{2}+(1-4Bt)^{2}\right] ^{1/2}.  \label{expectaionc}
\end{equation}

We will now turn our attention to the causal interpretation of the
wavefunctions $\psi ^{(I)},$ $\psi ^{(II)}$, and $\psi .$

\section{The causal interpretation}

The causal interpretation of the above minisuperspace models is
straightforward. Let us begin by the dust field. In this case, the quantum
trajectories are solutions of the following differential equation: 
\begin{equation}
p_{x}=\frac{1}{2}\dot{x}=\frac{\partial S_{d}}{\partial x},  \label{dust}
\end{equation}
where $S_{d}$ is the phase of the wave function (\ref{psic}). The general
solution is: 
\begin{equation}
x(t)=x_0\;[16b^{2}t^{2}+(1-4Bt)^{2}]^{1/2},  \label{sd0}
\end{equation}
where $x_0$ is an arbitrary positive integration constant. That the mean
value of $x$ founded in the previous section is, apart from the integration
constant, the same function of time as the solution (\ref{sd}) is not
surprising. Mean values in the causal interpretation are the same as in the
usual interpretation if we also assume that the amplitude squared of the
wave function, $R^{2}(x,0)$, represents a probability density distribution
of initial conditions of the quantum trajectories. Then, we obtain from Eq. (%
\ref{gauss2}) 
\begin{eqnarray}
\langle x(t)\rangle &=&\int_{0}^{\infty }\left( {\frac{8b}{\pi }}\right)
^{1/2}e^{-2bx_0^{2}}x_0[16b^{2}t^{2}+(1-4Bt)^{2}]^{1/2}dx_0  \nonumber
\label{sd} \\
&=&{(2\pi b)}^{-1/2}[16b^{2}t^{2}+(1-4Bt)^{2}]^{1/2} \; ,
\end{eqnarray}
which is the result (\ref{expectaionc}) of the previous section.

It can be seen from Eq. (\ref{sd0}) that no quantum trajectory is singular.
The scale factor is never zero for any $x_0$ greater than zero. Also, the
trajectories approach the classical one for large $|t|$ (remember that $%
x(t)\propto a^{3/2}(t)$). This behavior can be seen by examining the quantum
potential,
\begin{equation}
Q_{d}=-2b\frac{[2b(x^{2}-8bt^{2})-(1-4Bt)^{2}]}{[(1-4Bt)^{2}+16b^{2}t^{2}]^2}%
.  \label{qpd}
\end{equation}
Along the quantum trajectory (\ref{sd0}), the quantum potential turns out to
be: 
\begin{equation}
Q_{d}(t)=-\frac{2bx_0^{2}(1-2bx_0^2)}{x^2(t)},  \label{qpdt}
\end{equation}
where $x(t)$ is given by (\ref{sd0}).

From the above equation we can see that the quantum potential goes to zero
when $|t|$ becomes large but it is positive of order $b$ when $|t|$ is small%
\footnote{%
There is nothing special with the choice $x_0=1/\sqrt{2b}$. The quantum
potential is zero along the trajectory but the quantum force $%
F_d(t)=-\partial _x Q_d$ is not.}. In this situation, it works like a
repulsive force around the region $x=0$. Figure 1 shows the curves $x(t)$
and $Q(t)$ (along the trajectory) versus $t$. They represent classical
universes contracting from infinity to a minimum size, where its behavior is
not classical, and then expanding to infinity, getting classical again as
the scale factor becomes large. These models have no particle horizon. The
integral $\int _{-\infty} ^t a^{-1}(t^{\prime}) {\rm d}t^{\prime}\propto
\int _{-\infty} ^t x^{-2/3}(t^{\prime}) {\rm d}t^{\prime}$ diverges.
\INSERTFIG{4.2}{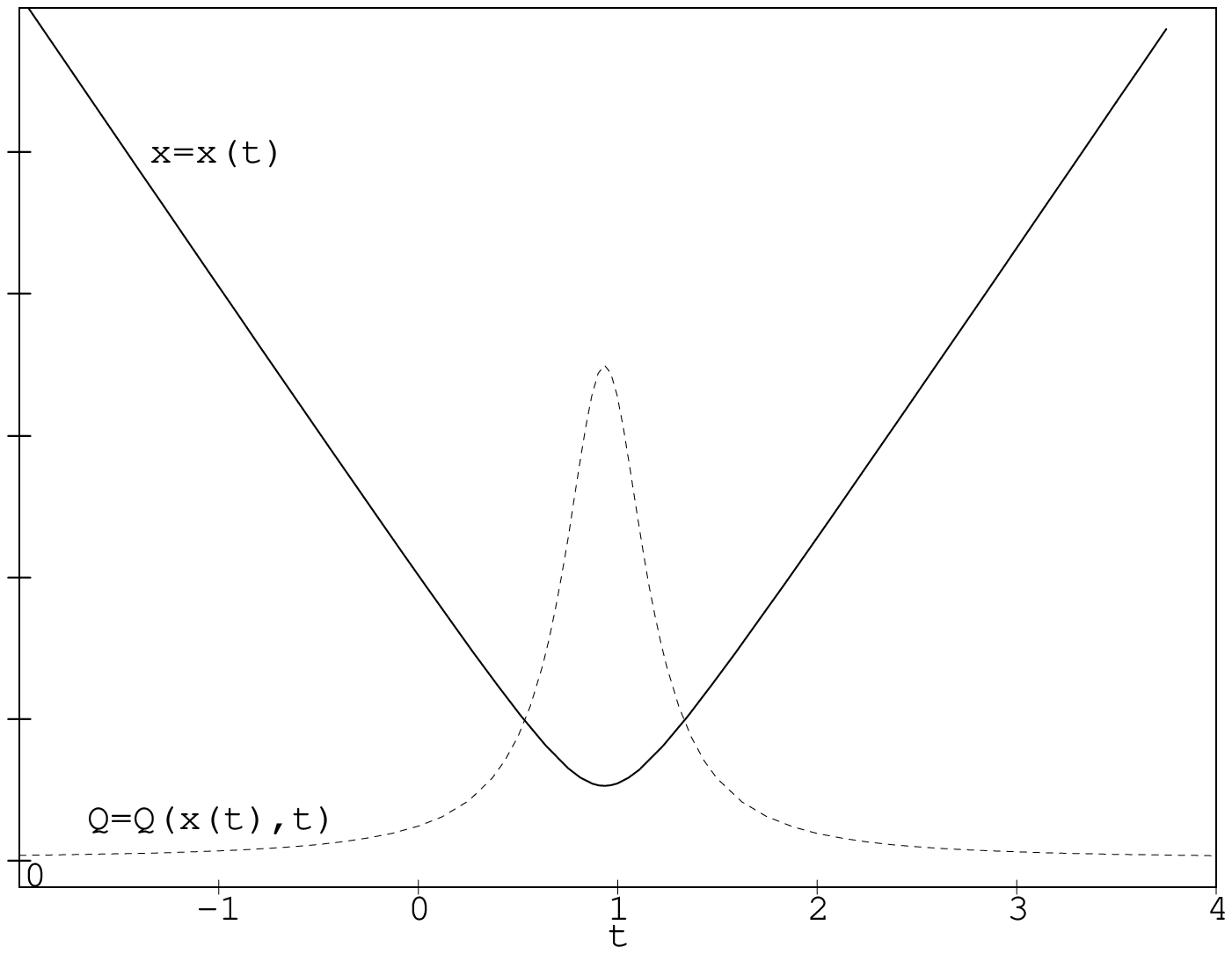}{Fig.~1: A typical trajectory for 
$x(t)$ and its corresponding quantum potential (dashed line). We see 
that when  $x$ approaches
the classical singularity the quantum potential becomes large.}

Let us now turn our attention to the more realistic case of radiation.
Again, the application of the causal interpretation is straightforward. We
have to take the phases of the wave functions (\ref{psia},\ref{psib}) for
the cases $k=0,k\pm 1$, calculate their derivatives with respect to $a$, and
equate the result with $p_{a}=12\dot{a}$. The solutions of these first order
differential equations are:

\begin{equation}
a(t)=a_0\left\{ 
\begin{array}{ll}
\sqrt{b^{2}\sin ^{2}t+(6-B\tan t)^{2}\cos ^{2}t} & \quad \mbox{for }k=1 \\ 
\sqrt{b^{2}t^{2}+(6-Bt)^{2}} & \quad \mbox{for }k=0 \\ 
\sqrt{b^{2}\sinh ^{2}t+(6-B\tanh t)^{2}\cosh ^{2}t} & \quad \mbox{for }k=-1
\end{array}
.\right.  \label{abohm}
\end{equation}
where $a_0$ is a positive integration constant. Like before, they have the
same functional behaviour as the mean values encountered in the previous
section for the reasons already explained in the dust case. None of these
solutions reach the singular point $a=0$ .The quantum potentials are
given by
\begin{equation}
Q = 3b\left\{ 
\begin{array}{ll}
\frac{[b(b\sin ^2(t)-72a^2)+(6-B\tan (t))^2 \cos ^2(t)]}{[(6-B\tan
(t))^2\cos ^2(t)+b^2\sin ^2(t)]^2} & \quad \mbox{for }k=1 \\ 
\frac{[b(bt^2-72a^2)+(6-Bt)^2]}{[(6-Bt)^2+b^2t^2]^2} & \quad \mbox{for }k=0
\\ 
\frac{[b(b\sinh ^2(t)-72a^2)+(6-B\tanh (t))^2 \cosh ^2(t)]}{ [(6-B\tanh
(t))^2\cosh ^2(t)+b^2\sinh ^2(t)]^2} & \quad \mbox{for }k=-1
\end{array}
.\right. 
\end{equation}
At the trajectories, the quantum potentials for
$k=0,1,-1$ are the same, and equal to
\begin{equation}  \label{qp1} 
Q = \frac{3ba_0^2(1-72ba_0^2)}{a^2(t)}, 
\end{equation}
where $a(t)$ is given by (\ref{abohm}).

For $k=0$ and $k=-1$ the models are qualitatively similar to the dust case.
It can be seen from Eq. (\ref{abohm}) that these universes contract
classically from infinity to a minimum size, where their behaviors are not
classical, and then expand to infinity, getting classical again as the scale
factor becomes large. The case $k=1$ deserves special attention. Examining
the case with $B=0$ and $b>6$ (the general case is qualitatively similar),
we can see that the quantum trajectory oscillates between the minimum value $%
6a_0$ and the maximum value $a_0 b$. Their ratio is $b/6$.
The ratio between the quantum and classical forces at these points are
$-b^2/36$ 
for the minimum and $-36/b^2$ for the maximum. The universe we live in is
large and classical, and it must have gone through a contracted phase where
nucleosynthesis took place. This means that the ratio between the maximum
and minimum values of the radius of the universe (which is the inverse ratio
of the respective temperatures at these epochs since the fluid is radiation)
must be at least of the order of $10^{10}$. Hence $b$ must be greater than $%
10^{10}$, yielding a very flat initial gaussian  wave function. This ensures that the
universe had a nucleosynthesis era with a classical behaviour since then.

These models do not have particle horizons. In this case the particle
horizon is proportional to the integral $\int _{-\infty} ^t
a^{-1}(t^{\prime}) N(t^{\prime}) {\rm d} t^{\prime}$ which in the gauge $N=a$
we are using evidently diverges.

Figures 2 and 3 show the scale factor and the quantum potential for $k=0$
and $k=-1$. When $a$ is large, the quantum potentials go to zero, while when 
$a$ approaches its minimum size, they become important, creating an
effective repulsive quantum force around this region. Figures 4 and 5 show
the scale factor and the quantum potential for $k=1$ and for different
values of $b$. Note that for larger $b$'s the amplitude of
oscillation also becomes larger.

\INSERTFIG{4.2}{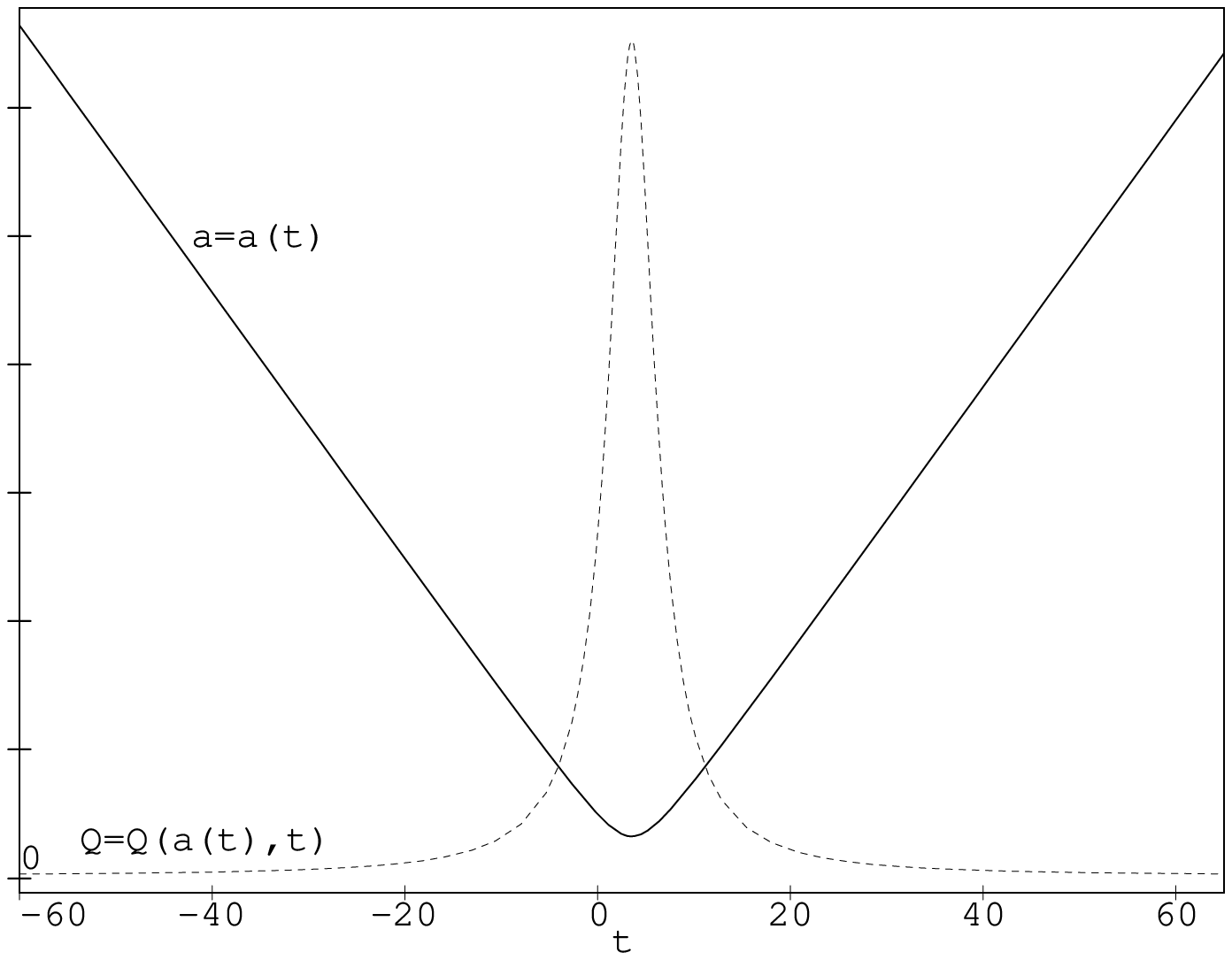}{Fig.~2: Typical quantum trajectory with its
corresponding quantum potential (in dashed lines) for the radiation
fluid model for the case where $k=0$.}

\INSERTFIG{4.2}{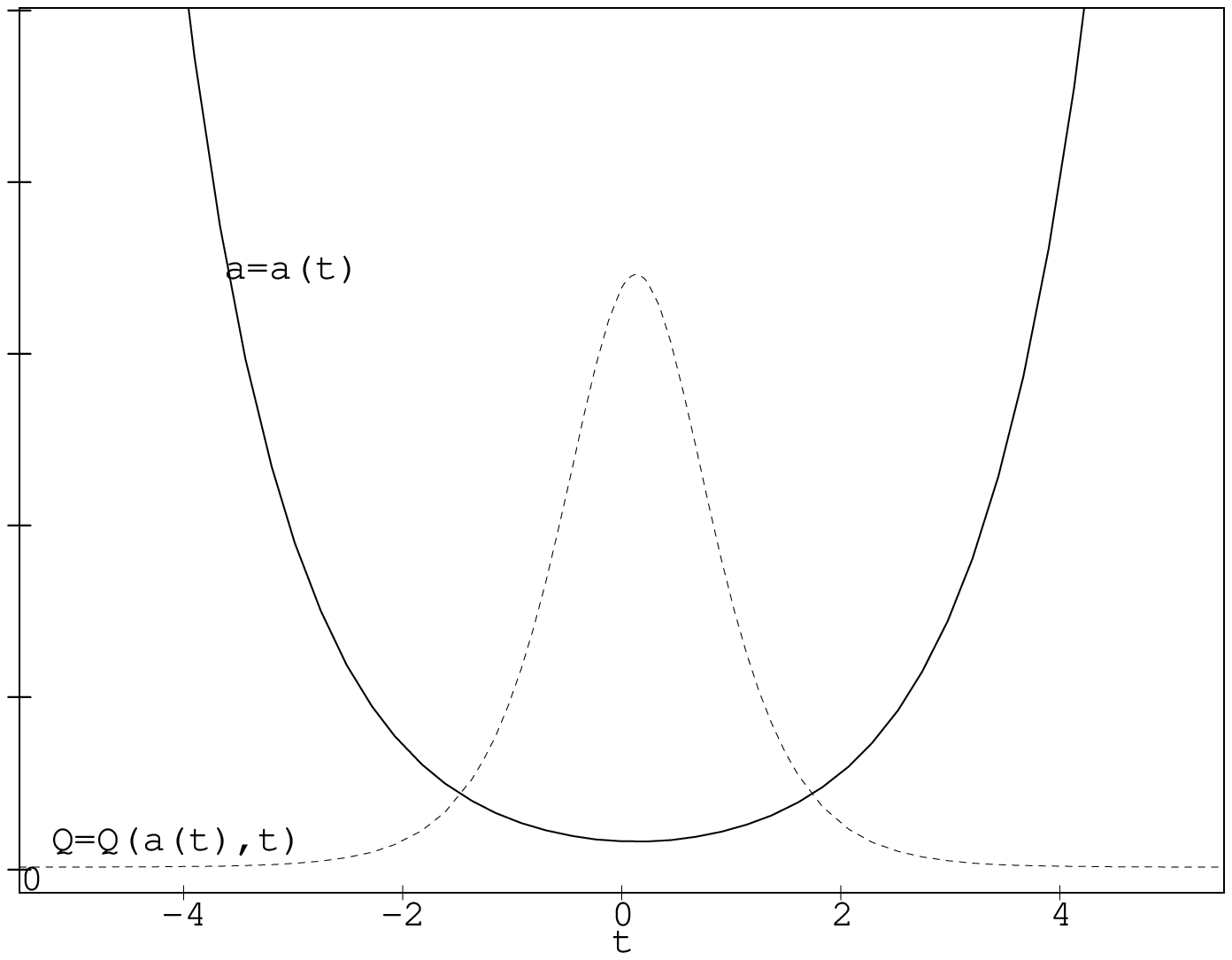}{Fig.~3: Typical quantum trajectory with its
corresponding quantum potential (in dashed lines) for the radiation
fluid model for the case where $k=-1$.}

\INSERTFIG{4.2}{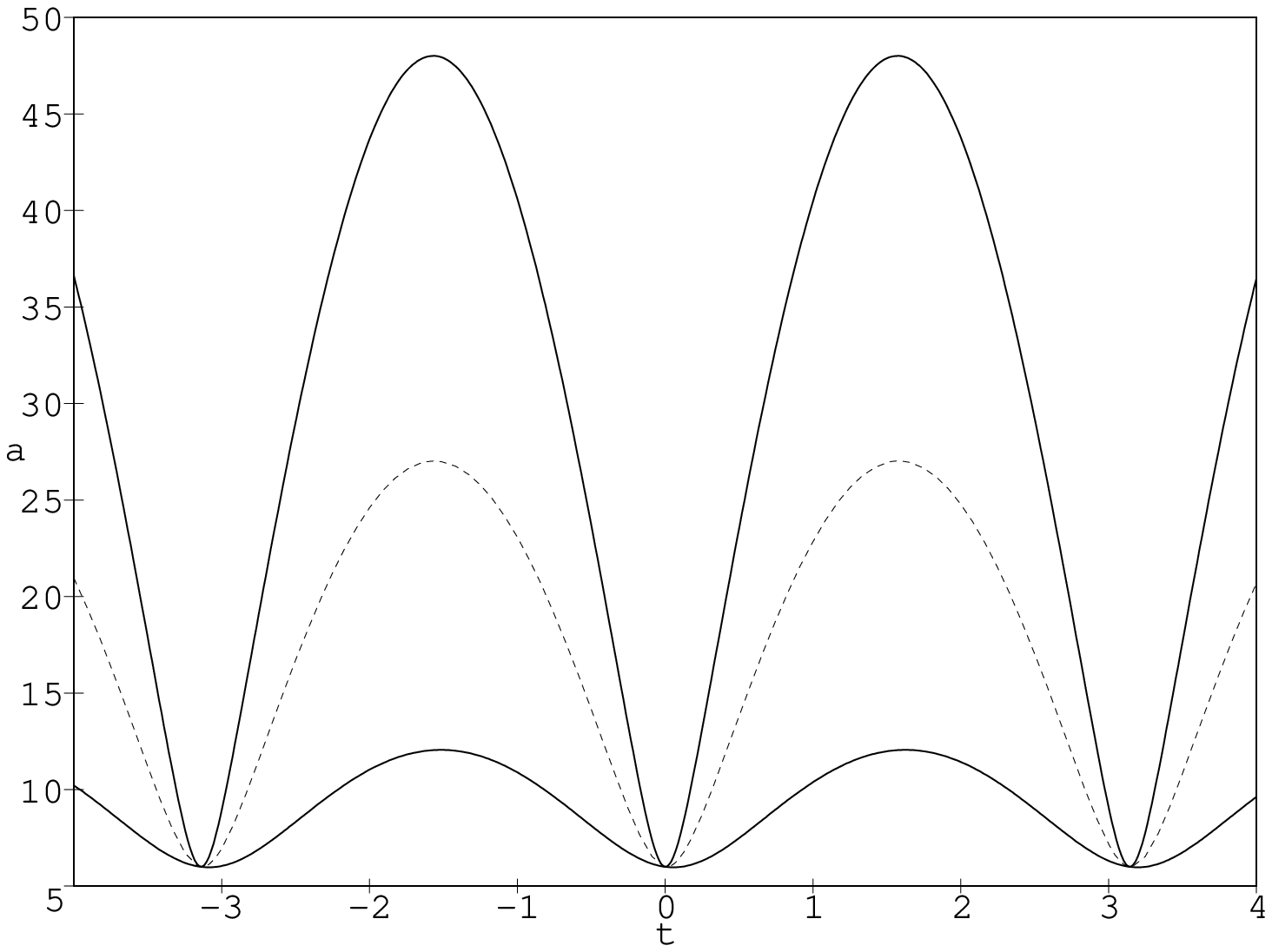}{Fig.~4: Quantum trajectories for different
values of $b$ and the same initial condition $a(0)$ for the radiation
fluid model and $k=1$. The larger the value of $b$
the larger is the amplitude of oscillation.}

\INSERTFIG{4.2}{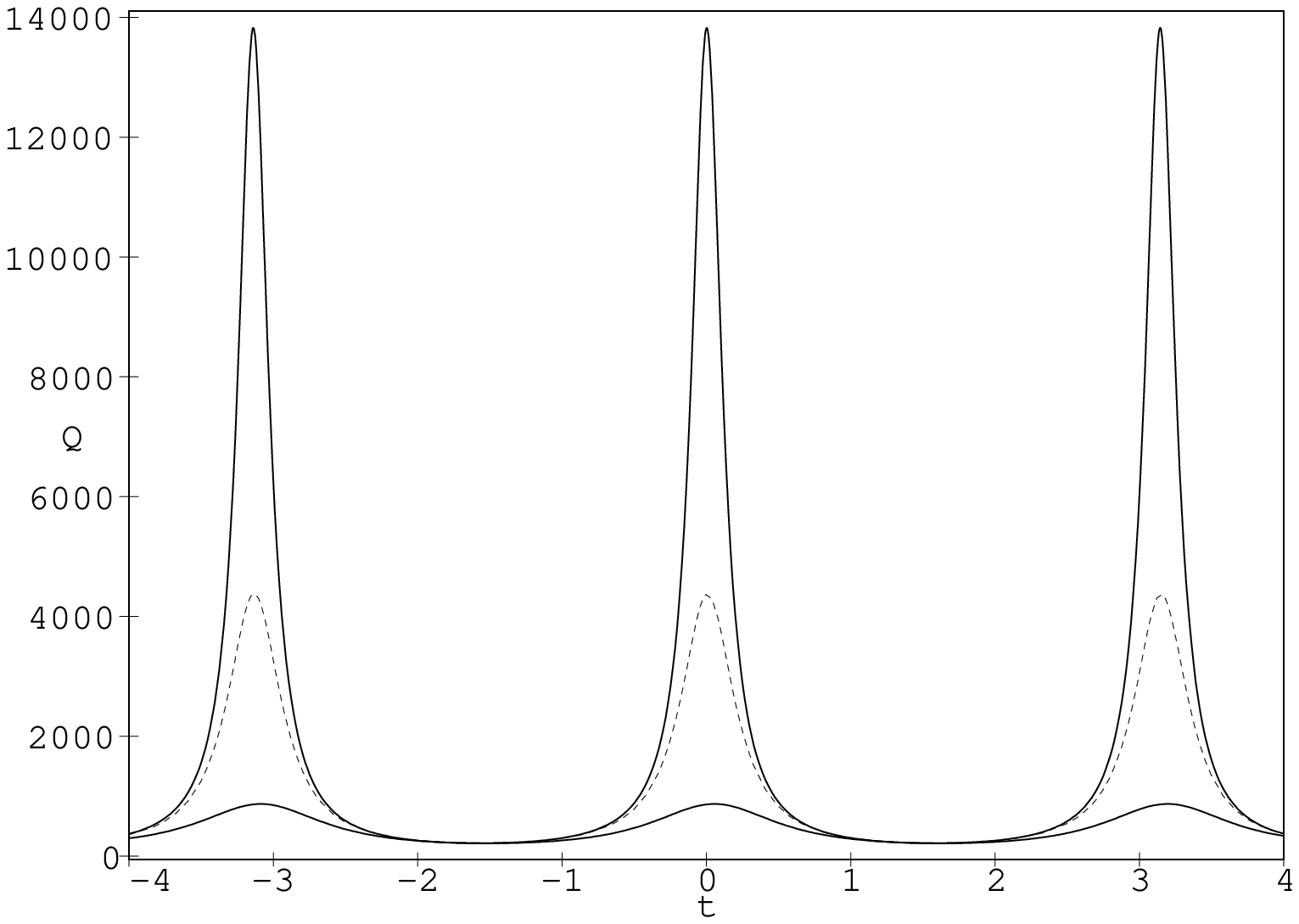}{Fig.~5: Quantum potentials computed along
the quantum trajectories shown in Fig.~4. The dashed quantum potential
corresponds to the dashed trajectory in Fig.~4. We can see that the
value of the quantum potential increases as we approach a classical
singularity, being higher when $b$ is higher.}

\section{Conclusion}

In this letter we have obtained a class of non-singular cosmological models
without particle horizons by applying the causal interpretation to some
quantum states of the universe already obtained in the literature \cite
{gotay,lemos} and presented in section 3. We have shown that the Bohmian
trajectories are the same functions of time as their corresponding
mean-values obtained in Refs. \cite{gotay,lemos}, which makes use of the
conventional Copenhaguen interpretation. However, it seems to us that only
in the causal approach we can arrive at definite conclusions about the
existence of singularities. It is not because the mean value of a variable
is different from zero at all times that this variable cannot be zero
sometime. Nothing forbids that this be the case for the single universe we
live in. The causal interpretation applied to this problem states that each
individual trajectory of these quantum states is not singular, a much more
stronger and valuable result. This means that if the universe is in one of
these quantum states, then it is indeed non-singular because there is no
quantum trajectory with singularities.

For the more realistic radiation fluid, we have obtained singularity-free
models, without particle horizons, which approach the classical behavior as
the universe expands. For flat and negative curvature three-surfaces this is
the situation in all cases. The universe contracts classically from infinity
to a minimum size, where its behavior is not classical, and then expands to
infinity, getting classical again as the scale factor becomes large. For the
positive curvature case, the classical behavior is achieved only for some
values of the parameters. In this case, we would have an eternal periodic
universe which is classical when it is large.

These examples show that quantum gravitational effects can indeed prevent
the formation of cosmological singularities but this result can only be
stated with strength along the lines of the causal interpretation. It should
be interesting to investigate if the results presented here are stable under
small perturbations. Then we will have to face new technical and
interpretational problems. But this is another story.

\section*{Acknowledgments}

Part of this work was done while NPN was a PREVI (Special Program for
Visiting Professor/Researcher) fellow at the Physics Department of the
Federal University at Juiz de Fora. We would like to thank the group of the
``Pequeno Semin\'ario" at CBPF for useful discussions and FAPEMIG for
financial support. NPN would like to thank CNPq for financial support and
the Federal University at Juiz de Fora, UFJF, for hospitality. JAB would
like to thank the Laboratory for Cosmology and Experimental High Energy
Physics (Lafex) at the Brazilian Center for Physical Research (CBPF/CNPq)
for hospitality. MASL is supported by a scholarship from Capes.

\end{document}